\UseRawInputEncoding
\documentclass[a4paper,fleqn,usenatbib,useAMS]{mnras}

\usepackage{graphicx}	
\usepackage{amssymb}	
\usepackage{bm}		
\usepackage{amsmath}
\usepackage{amssymb}
\usepackage{lscape} 
\usepackage{longtable}

\usepackage{float}

\usepackage[T1]{fontenc}
\usepackage{ae,aecompl}

\usepackage{txfonts}

\title{The Galaxy Population of the Core of Coma Cluster}

\author[Nagamani P., Hasan Priya and Hasan S N]{Nagamani P.$^{1}$\thanks{E-mail:nagamani.poloji@gmail.com} Hasan P.riya $^{2}$\thanks{E-mail: priya.hasan@gmail.com} 
and Hasan S. N. \thanks{E-mail:hasan.najam@gmail.com}$^{3}$
\\
$^{1}$Department of Astronomy, Osmania University, Hyderabad, India\\
$^{2}$Department of Physics, Maulana Azad National Urdu University, Hyderabad, India\\
$^{3}$Department of Mathematics, Maulana Azad National Urdu University, Hyderabad, India
}

\date{Accepted 2021 November 22. Received 2021 November 22; in original form 2021 March 08}

\pubyear{2020}

\begin{document}
\pagerange{\pageref{firstpage}--\pageref{lastpage}}
\maketitle
\begin{abstract}
In this paper we present the structural properties and morphology of galaxies in the central region of the Coma Cluster brighter than $19.5^m$ in the $F814W$ band.  from the HST/ACS Coma Cluster Treasury Survey. Using mainly spectroscopic redshifts, we find 132 members from our sample of 219 galaxies.  In our sample of 132 members, we find 51 non-dwarfs and 81 dwarfs  and amongst our 32 non-members, we find 4 dwarfs and 28 non dwarfs. We do not have redshifts for the remaining 55  galaxies. We present bulge-disc decomposition of the sample using GALFIT and obtain parameters for our sample. Using  visual inspection of residuals, we do a  a morphological classification of the galaxies.  We studied the relation of morphological types with Bulge to Total Light Ratio ($B/T$), color magnitude relation (CMR), S\'ersic index ($n$), Kormendy relation and  cross-correlations between these parameters for the bulges and galaxies.  This work helps us understand important relations between various parameters like $B/T$, color and $n$ as well as insights into the  merger history of these galaxies in terms of their positions in the Kormendy Diagram and their S\'ersic  indices. Using statistical methods, we find that the there are significantly more E/SO, ESOs galaxies in the member population compared to non-members.

\end{abstract}





\section{Introduction}
The morphology of a galaxy is the result  of galaxy formation and evolution, interaction with the environment, internal perturbations and dark matter interactions \citep {1975gaun.book.....S, 2013seg..book..155B}. Visual inspection is the primary method to classify galaxies, for example, in the Galaxy Zoo project \citep{2008MNRAS.389.1179L}. To quantify morphology, structural properties of galaxies are found  by fitting radial profiles like S\'ersic, Nuker, Gaussian, King etc. to the luminosity profiles of galaxies.
		
 The morphology-density relation \citep{1980ApJ...236..351D, 1998gaas.book.....B} suggests that the number of elliptical galaxies increases with the richness of the galaxy cluster. \cite{1931ApJ....74...43H, 2000ApJ...542..673F, 2003MNRAS.339L..29H, 2009APh....31..255D} found that early-type galaxies are more abundant in clusters while late-type galaxies are more prevalent in the field. Correlations between morphology and structural parameters like effective radius ($r_{e}$), disk-scale length ($r_{s})$,  bulge to total light ratio ($B/T$) and  S\'ersic index ($n$) have been studied \citep{2004ApJ...602..664G,2011MNRAS.411.2439H,2014MNRAS.441.3083W,2012ApJ...746..136M}. Some studies have shown that high-luminosity galaxies are preferably early-type galaxies (E/SO)\footnote{Morphological types: central Dominant (cD) Elliptical/Lenticular (E/SO), Elliptical/Barred Lenticular (E/SBO), Lenticular (SO),  Barred Lenticular (SBO), Spirals (Sp), Barred Spirals (SBp), Irregular (Irr).} \citep{2003AAS...20314501B,2013seg..book..155B}.
 Morphological transformations can be driven by mergers both major and minor and/or gravitational interactions with neighbours. 
 
 Galaxy clusters are one of the largest structures in the universe which consists of hundreds to thousands of gravitationally bound galaxies with typical masses ranging from $10^{14}\--10^{15} M_{\odot}$.  There are three types of galaxy clusters: cD, spiral-rich and spiral-poor \citep{1974ApJ...194....1O}. Clusters with cD galaxies are rich in ellipticals, spiral-rich clusters have compositions similar to that of the field and spiral-poor clusters are dominated by SO galaxies. The above paper suggested that spiral-poor clusters represent a later evolutionary state of spiral-rich clusters, but cD clusters are intrinsically different. 

The Coma Cluster  is a  very rich and dense cluster at a redshift of 0.023 which corresponds to an approximate distance of 100 Mpc \citep{2010ApJS..191..143H}. This is  one of the most suitable clusters to study luminosity, environment and morphological  classification of galaxies. Coma is a cD type  cluster whose E:S0:Sp ratio is 3:4:2 \citep{1974ApJ...194....1O}. Extensive work has been done on morphological studies of galaxies in  the  Coma Cluster  \citep{1996A&A...314..763A, 1997MNRAS.285L..41G, 1998ApJ...500..750K, 2000A&AS..141..449M, 2002ApJS..138..265K, 2004ApJ...602..664G, refId0, 2014MNRAS.441.3083W, 2012ApJ...746..136M, 2015MNRAS.453.3729H}.

\cite{2004ApJ...602..664G} used data from the Issac Newton Telescope (INT) and studied galaxy properties using the IRAF ellipse task.  \cite{2004AJ....127.1344A} classified  galaxies based on $B/T$ parameter for INT data by using Monte Carlo simulations to get structural properties.  \cite{2011MNRAS.411.2439H} and \cite{2012ApJ...746..136M}  have used HST-ACS  data. \cite{2011MNRAS.411.2439H} made a comparison of the results of GALFIT and GIM2D for the complete sample using single S\'ersic fits without considering membership criteria but clearly indicated GALFIT as preferable. \cite{2012ApJ...746..136M} did bulge-disk and bar decomposition of galaxies using GALFIT only for members. 

In this paper, we make a detailed study of galaxy morphology using visual inspection of the residuals from  single S\'ersic profiles as well as bulge-disk (S\'ersic + exponential) decomposition. We  compare the structural and photometric  properties of the disc and  bulge of  the member and the non-member galaxy population in the central core of the Coma Cluster (within 0.5 Mpc ~$\approx 0.3^0$).  Updated membership using more recent spectroscopic data has been used in this analysis. We also make a detailed study of the morphological parameters $B/T$,  S\'ersic index $n$, color, surface brightness and their correlations for our sample.  This kind of analysis of HST-ACS data has not been done by previous authors. 
These results give us a deeper understanding of galaxy origin, evolution, merger history and their dependence on environment  in the core of Coma Cluster.

The paper is structured as follows: Section~1 of the paper gives a brief introduction and a description of the problem. Section~2 describes the HST-ACS observations. Section~3 describes the Membership data and sample selection. Section~4 describes the two dimensional decomposition and the GALFIT results used to get structural properties of galaxies by fitting functions to the luminosity profiles. Section~5 describes the results and discussions and Section~6 presents the conclusions of our paper.  

\section{Observations}
The Coma Cluster is one of the nearest rich clusters of galaxies and was the target of a deep two passband $F814W \approx 8146.9$ \AA \ and $F475W \approx 4794.0$ \AA \  imaging survey of the HST-ACS Treasury program \citep{2008ApJS..176..424C}.

\begin{figure}
\centering
\includegraphics[width=8cm,height=7cm]{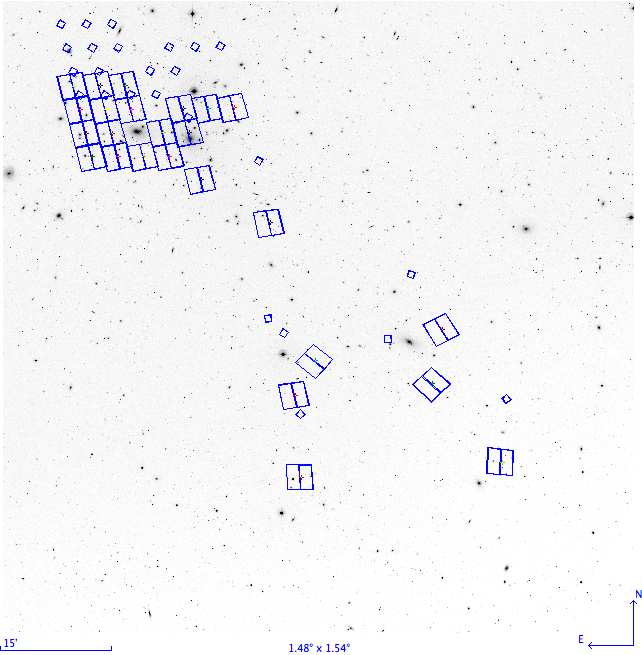}
\caption{HST-ACS Survey of Coma Cluster:  Tiles shown have some or all of the proposed observations. The survey was incomplete due to the ACS failure in January 2007  \citep{2008ApJS..176..424C}.}
\label{clusplot}
\end{figure}

We have used publicly available data from the HST-ACS Coma Cluster Treasury Survey data release 2.1 \citep{2008ApJS..176..424C}.  The survey was originally designed to cover an area of 740 arcmin$^2$ in regions of varying density of both galaxies and intergalactic medium within the cluster (Fig. \ref{clusplot}). However, due to an ACS failure on January 27$^{th}$ 2007, the survey is only 28\% complete, with 21 ACS pointings (230 arcmin$^2$) and partial data for a further 4 pointings (44 arcmin$^2$). Nineteen tiles out of twenty-five are located within 0.5~Mpc  and the remaining 6 tiles are located between 0.9~Mpc and 1.75~Mpc southwest of the center of Coma Cluster. NGC~4874 is the one of the cD galaxies located near the center of the cluster and is in our sample. NGC~4889 is the other cD galaxy in the cluster, but is absent in our sample as its observations were not done. The predicted survey depth for $10 \sigma$ detections for optimal photometry of point sources is $26.8^m$ in $F814W$ filter and $27.6^m$ in the $F475W$ filter.

For extended sources, the predicted $10\sigma$ limits for a 1 arcsec$^2$ region are $25.0^m$~arcsec$^{-2}$ in $F814W$  and $25.8^m$ ~arcsec$^{-2}$ in $F475W$ .

The SourceExtractor catalogs (SExtractor version 2.5; \cite{1996A&AS..117..393B}) for the  25 fields in $F814W$ and $F475W$ passbands and the science images are available at archive.stsci.edu/prepds/coma/datalist2.1.html. We have used data for the $F814W$ band, which ensures the deepest data for structural and luminosity function studies \citep{2008ApJS..176..424C}.

\section{Membership and Sample Selection}
We have used published spectroscopic and photometric  data to determine membership of galaxies in our sample. Spectroscopic members were selected within the redshift range  $0.023 \pm 0.009$ \citep{2001ApJS..137..279M}.

The number of members from various sources for which redshifts are known are: 
SDSS DR7 \cite{2009yCat.2294....0A} (72), SDSS DR12 \cite{Alam_2015} (78), \cite{2001ApJS..137..279M} (68), \cite{refId0}(49), (\cite{2011MNRAS.412.1098M, 2010yCat..74041745M}) (33), \cite{2011ApJ...737...86C} (9), 
(\cite{2014MNRAS.440.1690H, 2015MNRAS.453.3729H}) (59), and \cite{2011MNRAS.414.3052D} (112) (the number in  round braces shows the number matched with our sample.
It may be noted that redshifts for single galaxies were obtained from more than one source. In the case of photometric redshifts, \cite{refId0} was used as it agrees best with spectroscopic values. We have used photometric redshift only for one member galaxy and 13
non-members in our sample. 
  
\begin{figure}
\centering
\includegraphics[width=7cm,height=3cm]{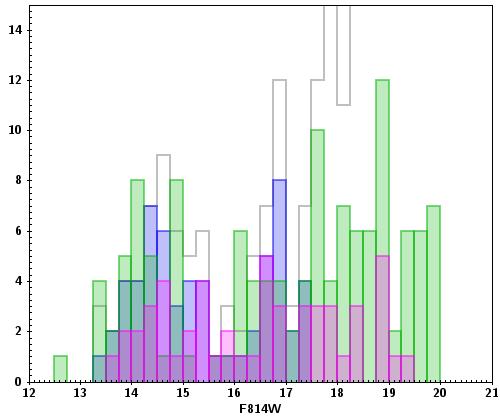}\\
\caption{The (F814W) magnitude  distribution of available spectroscopic data is shown in the figure. Here The grey steps represents the HST-ACS Coma Cluster SExtractor data \protect\citep{2010ApJS..191..143H}, pink represents the data points of \protect\cite{2001ApJS..137..279M}, blue represents the data points of \protect\cite{Alam_2015} and green represents the data points of \protect\cite{2011MNRAS.414.3052D}. }
\label{spec}
\end{figure}

 In Fig. \ref{spec}, we see the available spectroscopic data as a function of magnitude and that  spectroscopy can be determined  for galaxies brighter than $F814W~19.5^m$ \citep{2001ApJS..137..279M} and $F814W~20.75^m$ \cite{2014MNRAS.445.2385D}. We have used an limiting magnitude error of $0.01^m$ and hence $19.5^m$ was set as the magnitude limit for our study.
The SExtractor source catalog  by \cite{2010ApJS..191..143H} has  a total of 77,787 objects. In the  catalog \cite{2010ApJS..191..143H}, the column called FLAG\_OBJ defines the nature of the source where  0-galaxy, 1-star, 2-cosmic rays and 3-image artifacts. We found 594 sources brighter than $19.5^m$ in the $F814W$ band, of which 219 are galaxies (excluding half-images) with  FLAG\_OBJ value zero.

Of the 219 galaxies,  132 are members, 32 are non-members and 55 galaxies do not have redshift values. For the 132 cluster members, we have used spectroscopic redshifts for  131 galaxies and photometric redshift for only one galaxy\footnote{This was a bright galaxy for which decomposition results seemed reliable.  \citep{refId0}}. For 32 non-members, 19 was on the basis of spectroscopic redshift and 13 was on the basis of photometric redshifts \citep{refId0}.

\section{GALFIT and Structural Decomposition}

We have done a single S\'ersic as well as a two-dimensional bulge disk decomposition using GALFIT \citep{2002AJ....124..266P} for 219 objects.  We aim to study the bulge and disk properties of the galaxy sample and hence fit the  radial profile functions `S\'ersic' and `Exponential' for bulge and disk respectively \citep{2002AJ....124..266P}.

S\'ersic profiles are described as,   
        $$ \Sigma (r)=\Sigma_{e} e^{-\kappa [(r/r_e)^{1/n}-1]}$$
        
Where $r_{e}$ is the effective radius of galaxy such that half of the total flux lies within $r_{e}$. $\Sigma_{e}$ is the pixel surface brightness at $r_{e}$ and $n$ is the S\'ersic index. The relationship between $r_{s}$ and effective radius $r_{e}$ of a S\'ersic profile is $$r_{e}=1.678r_{s}$$ where $r_{s}$ is the scale length. $r_{s}$ is defined as the radius at which intensity drops by $e^{-1}$.

Exponential disc profiles are given by,
 $$\Sigma (r)=\Sigma_{o}\ e^{(-r/r_{s})}$$
 
where  $\Sigma_{o}$ is the central surface brightness.

For the S\'ersic fit, the index $n$ is a free parameter and for the disk,  we fix  $n=1$. 

To run GALFIT, we have used input values from the SExtractor catalogues \cite{2010ApJS..191..143H}. These include input parameters like position, apparent magnitude, effective radius, $b/a$ and the positional angle. The input science images were extracted using IRAF and included the galaxy and a portion of the sky required for fitting. 
For the fitting region,  $xmin, xmax, ymin ymax$ were  taken from the  image header.
Tinytim \citep{2011SPIE.8127E..0JK} was used to create the point-spread function (PSF) image.  Masking was done for objects in the frame that did not belong to the galaxy using IRAF.

\begin{figure}
\begin{tabular}{ccc}
\centering
\includegraphics[width=2.5cm,height=2.5cm]{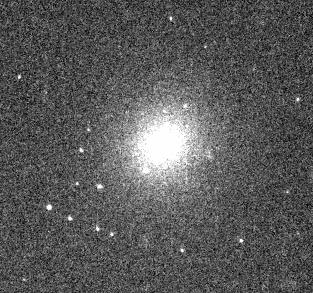}&\includegraphics[width=2.5cm,height=2.5cm]{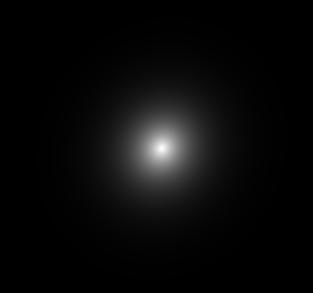}&\includegraphics[width=2.5cm,height=2.5cm]{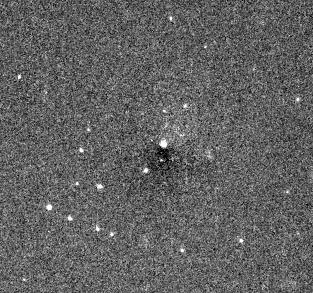}\\
\includegraphics[width=2.5cm,height=2.5cm]{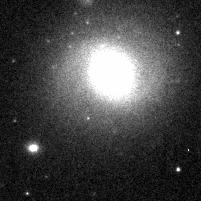}&\includegraphics[width=2.5cm,height=2.5cm]{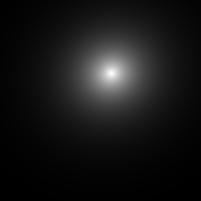}&\includegraphics[width=2.5cm,height=2.5cm]{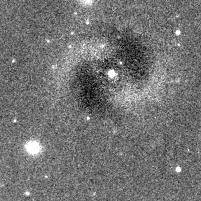}\\
\includegraphics[width=2.5cm,height=2.5cm]{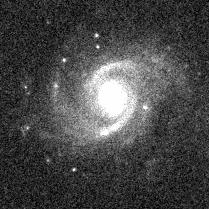}&\includegraphics[width=2.5cm,height=2.5cm]{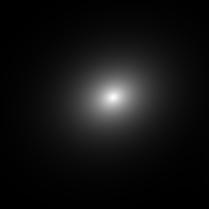}&\includegraphics[width=2.5cm,height=2.5cm]{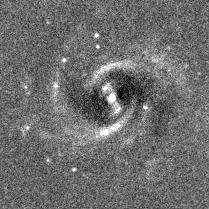}\\
\includegraphics[width=2.5cm,height=2.5cm]{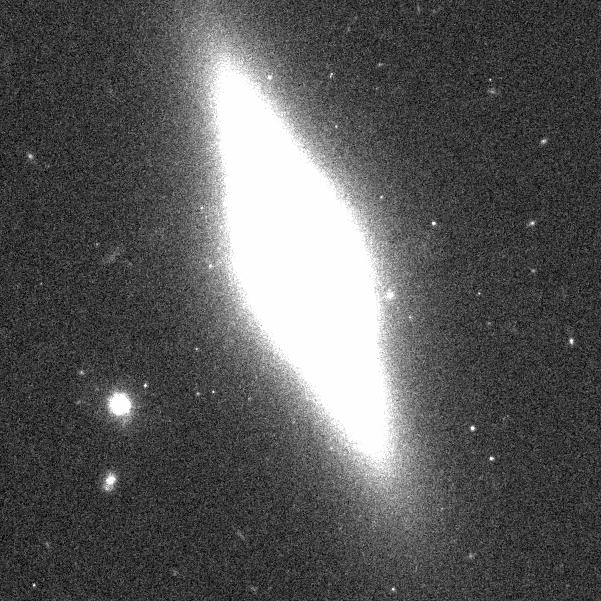}&\includegraphics[width=2.5cm,height=2.5cm]{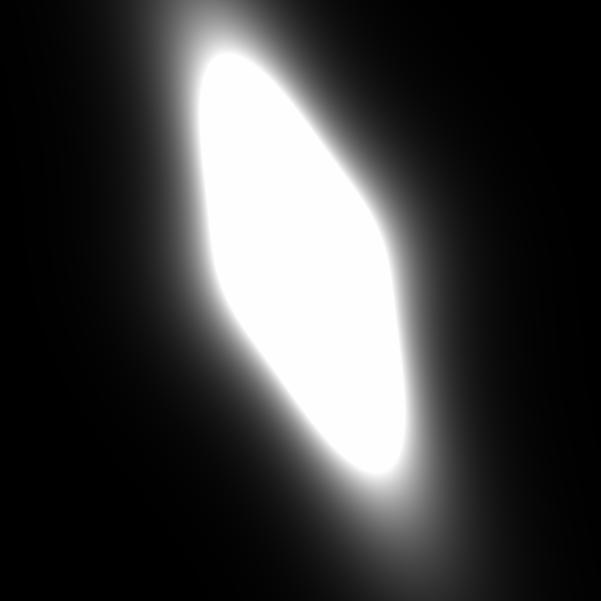}&\includegraphics[width=2.5cm,height=2.5cm]{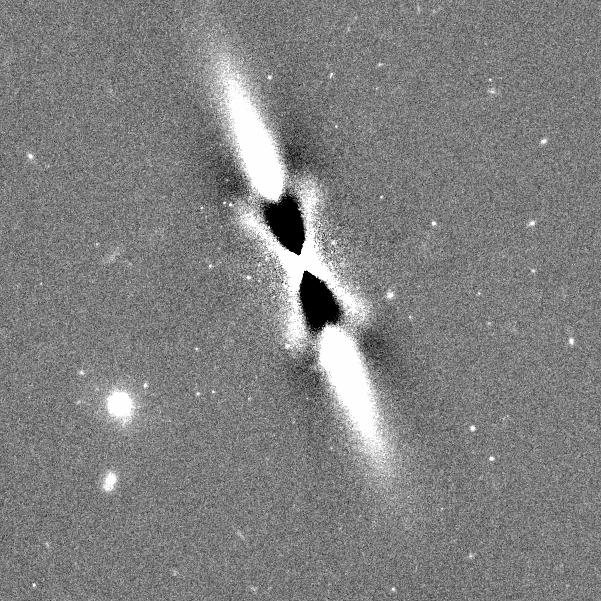}\\
\includegraphics[width=2.5cm,height=2.5cm]{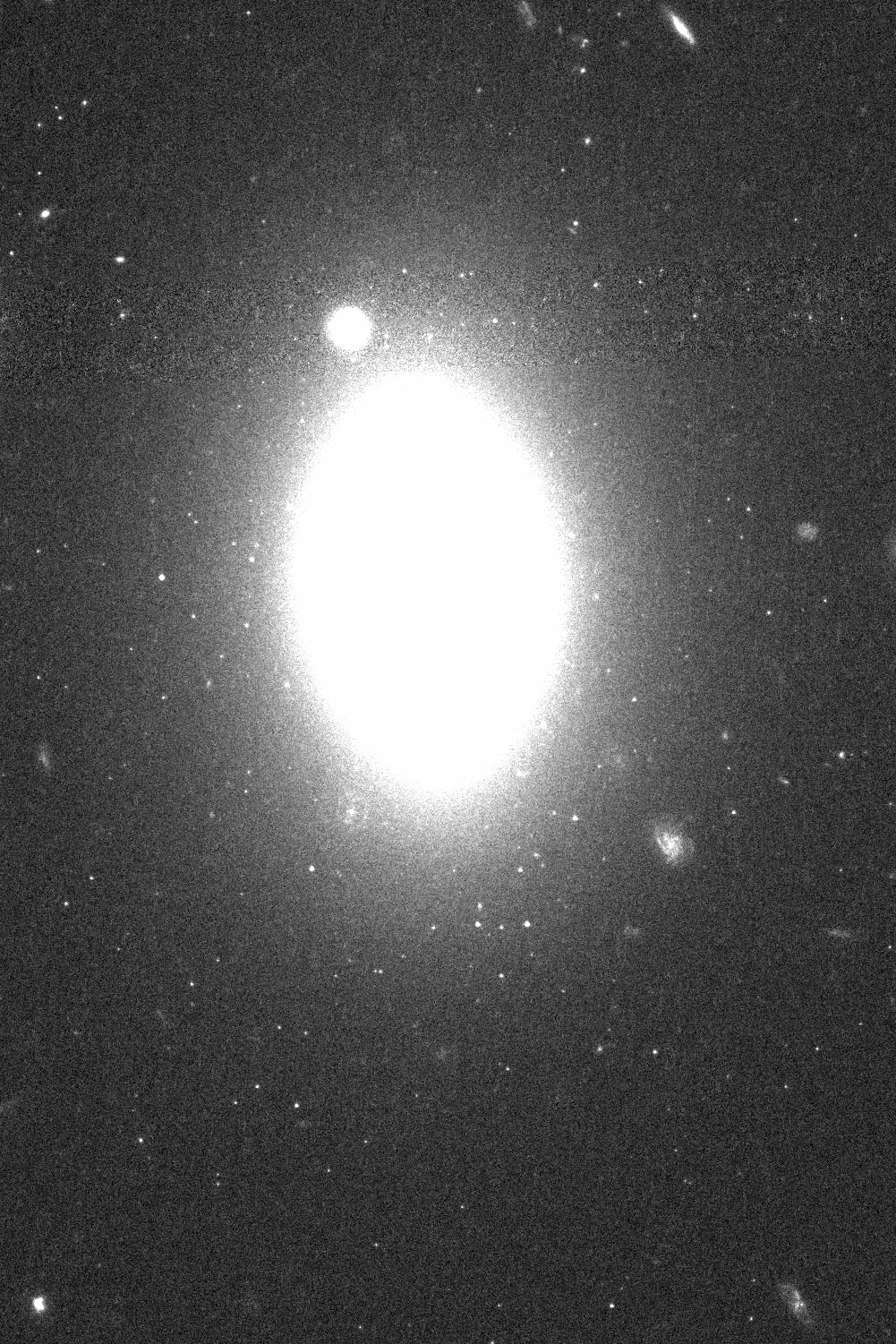}&\includegraphics[width=2.5cm,height=2.5cm]{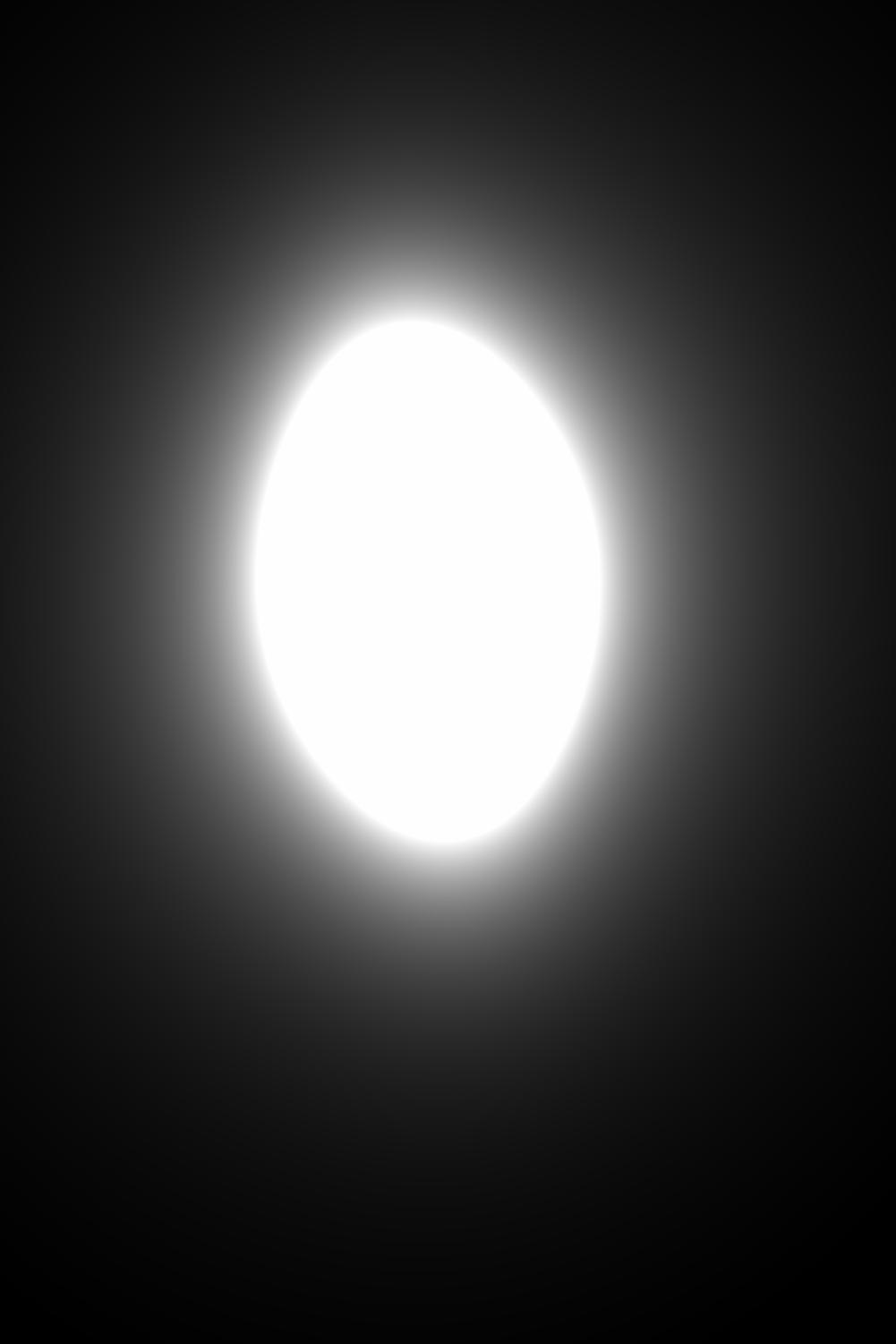}&\includegraphics[width=2.5cm,height=2.5cm]{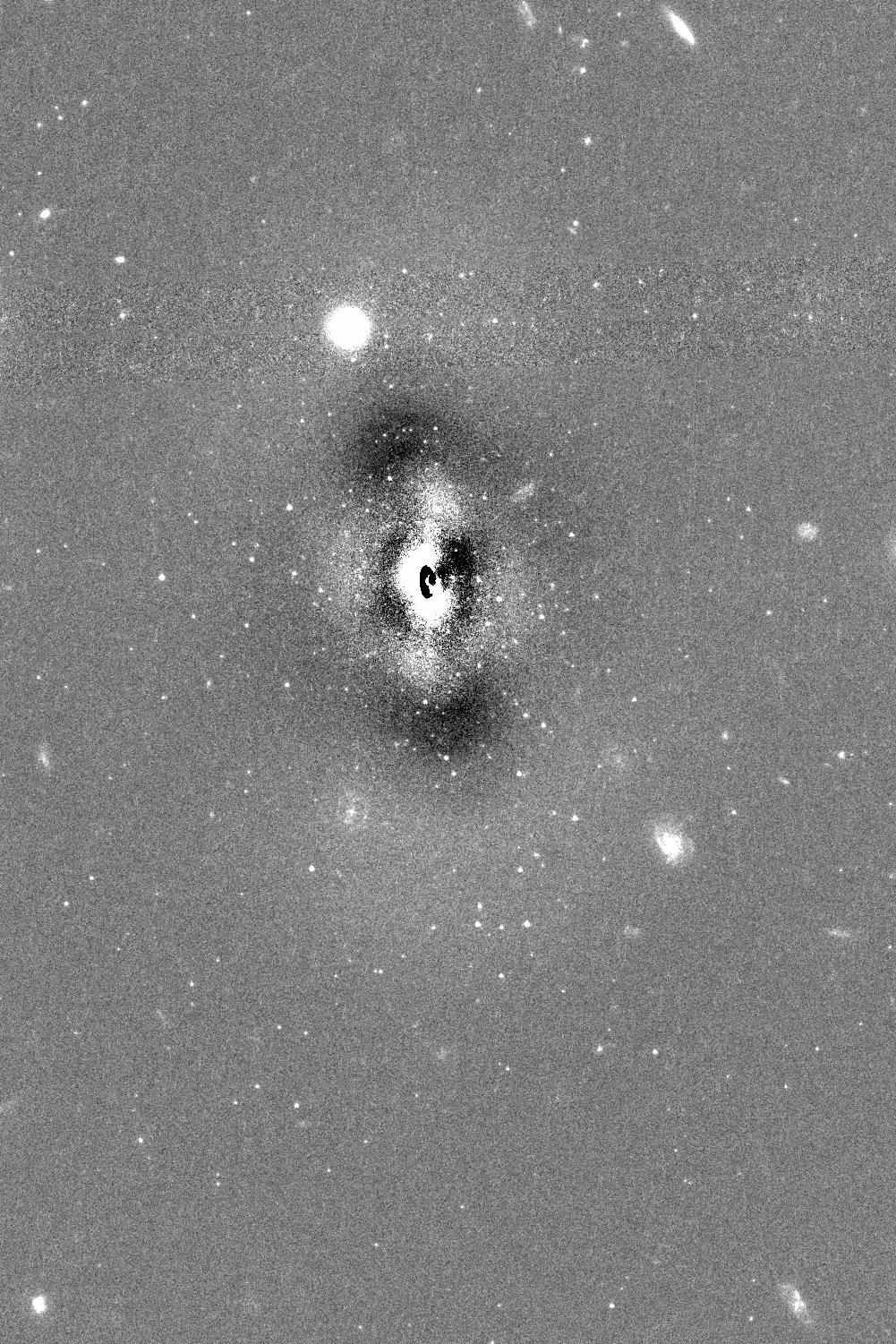}\\
\includegraphics[width=2.5cm,height=2.5cm]{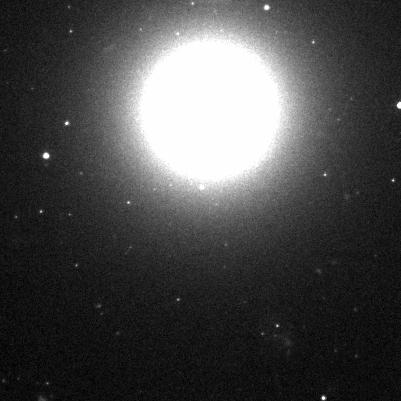}&\includegraphics[width=2.5cm,height=2.5cm]{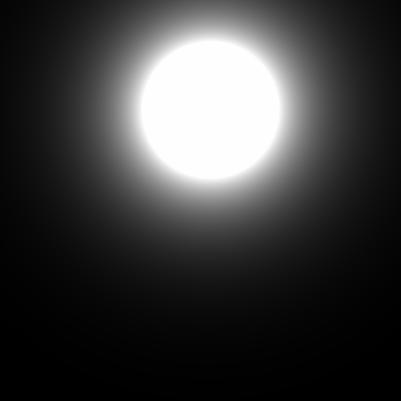}&\includegraphics[width=2.5cm,height=2.5cm]{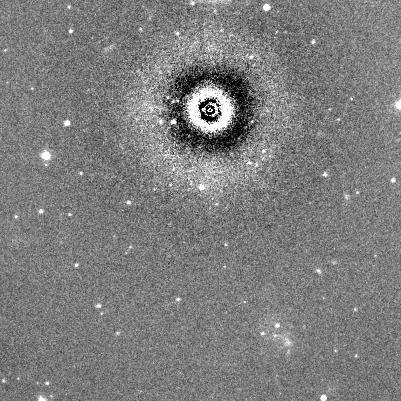}\\
\includegraphics[width=2.5cm,height=2.5cm]{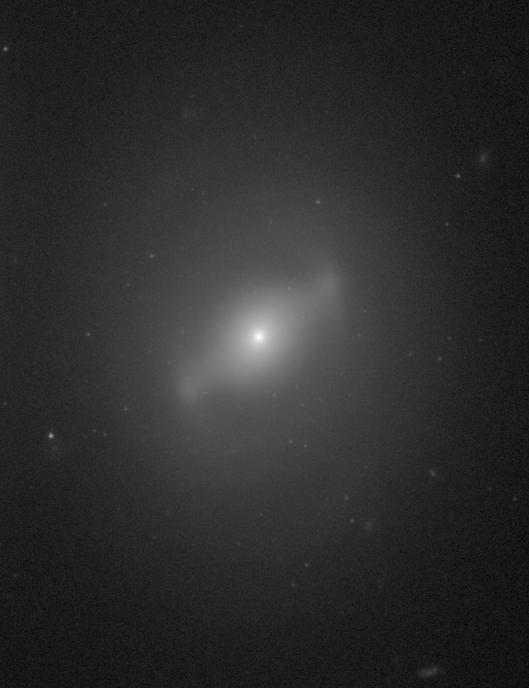}&\includegraphics[width=2.5cm,height=2.5cm]{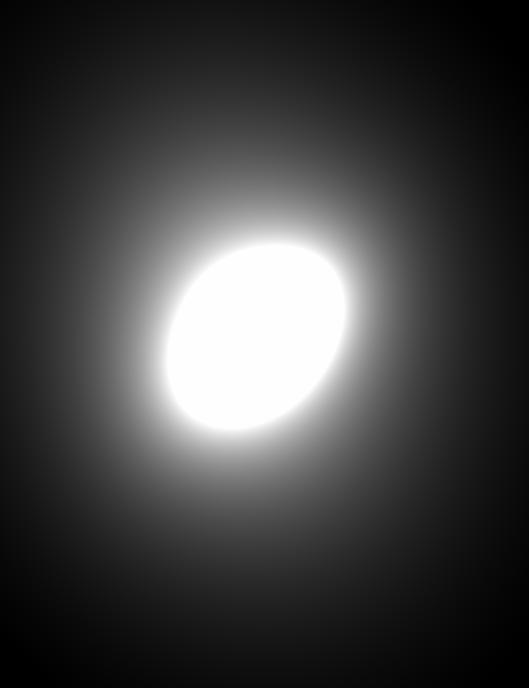}&\includegraphics[width=2.5cm,height=2.5cm]{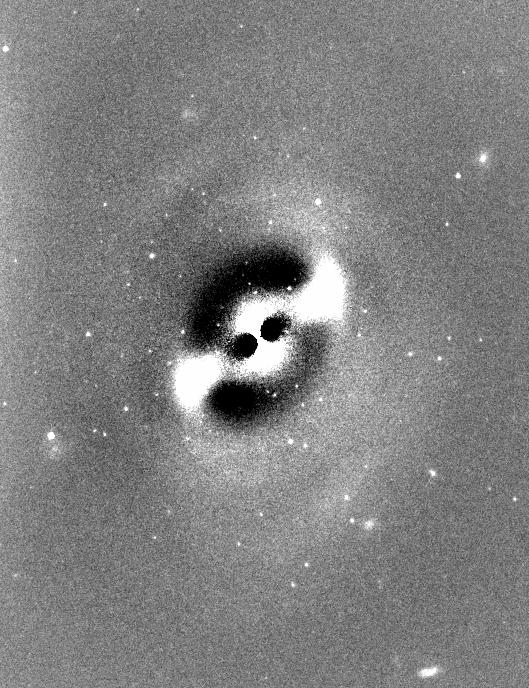}\\

\end{tabular}

\caption{The Figure shows  sample results of bulge disc decomposition by GALFIT. The first column shows HST-ACS science images, middle column shows GALFIT model images and the third column shows residuals which are obtained by subtracting the model image from the science image.
The  galaxy in the the row  has a zero residual and is classified as elliptical (E), second row is a spiral (Sp), third row is a barred spiral (SBp), fourth row is boxy/peanut bulge (E/SBO), fifth is E/SO, sixth is ring and last is SBO.}
\label{galfit}
\end{figure}

Figure \ref{galfit} shows a few examples of our GALFIT results:  The science image, model image and residual. The residuals are very useful to identify features of the galaxy and have been used to classify galaxies visually. 
Table \ref{galfitall} and Table \ref{terr}1  describes the column data from  detailed analysis and the errors  for our sample respectively.  The complete tables will be  available on request.
 
Tables \ref{singleser}2  \ref{bdc}3  describe the  comparison of results  of our sample with \cite{2011MNRAS.411.2439H,2014MNRAS.441.3083W,2014MNRAS.440.1690H}. We found a good agreement with \cite{2011MNRAS.411.2439H,2014MNRAS.441.3083W} for single S\'ersic fits and  found a good agreement with \cite{2014MNRAS.440.1690H} for bulge-disk decomposition fits. We also have added a comparison with  \cite{2014MNRAS.441.3083W} (multiple S\'ersic fits).

\begin{figure}
\centering
\includegraphics[width=7cm,height=4cm]{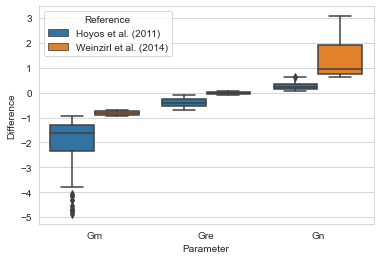}
\caption{Comparison of our data with  for single Sersic fits. }
\label{comp1}
\end{figure}

\begin{figure}
\centering
\includegraphics[width=7cm,height=4cm]{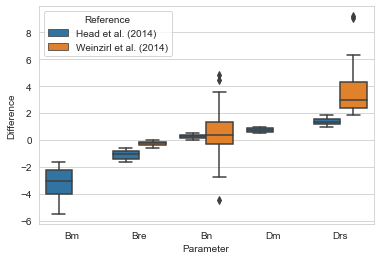}
\caption{Comparison of our data  for bulge-disk and multiple Sersic fits. }
\label{comp2}
\end{figure}

Figure \ref{comp1} and \ref{comp2} show boxplots of the comparison of our data with \cite{2011MNRAS.411.2439H,2014MNRAS.441.3083W} for single S\'ersic fits and  with \cite{2014MNRAS.440.1690H} (bulge-disk decomposition)  and \cite{2014MNRAS.441.3083W}  (multiple S\'ersic fits) respectively. 

\section{Results and Discussion }
\subsection{Morphology}

In this section, we describe the morphological classification of our sample galaxies. We divided our sample into cluster members, non-members and unknown redshift galaxies by taking the spectroscopic redshifts described in the earlier section. 

Visually,  we classified the galaxies based on  features in GALFIT residual images (Fig. \ref{galfit}) as GALFIT residuals show features like spiral arms, bars, lenticulars with and without bars and also by changing contrasts, etc. The morphological types were given based on Hubble's scheme as elliptical (E) lenticular (SO), lenticular with bar (SB0), spirals (Sp) and Barred spirals (SBp). When the residual images were not clearly distinguishable between ellipticals and lenticulars, we classified those galaxies as E/SOs. Our sample comprises of E/SO, E/SBO, SO, SBO, Sp, SBp, Ring and Irr galaxies.  \cite{2012arXiv1208.2295G} classified galaxies with boxy/peanut bulges and thick boxy bulges  as barred galaxies.  We followed the same scheme to  classify E/SBOs. The detailed distribution in shown in Table~\ref{morph}. Also, we have classified dwarfs and non-dwarfs for members and non-members of the cluster  based on the absolute  magnitude of the galaxies $F814W <-18.5^m$  \citep{2012ApJ...746..136M} and  using  $H_0$=70 km/s/Mpc, $\Omega_m$=0.3 and $\Omega_L$=0.7 \citep{2008ApJS..176..424C}. To calculate absolute magnitude we have taken the distance of 100 Mpc  and the distance modulus is 35 \citep{2008ApJS..176..424C, 2012ApJ...746..136M}. We found that of the 132 members, 51 are non-dwarfs and 81 are dwarfs. For the 32 non-members,  4 are dwarfs and 28 are non-dwarfs. For non-members, we have used  individual redshift values to calculate absolute magnitude. For the remaining 55 galaxies, we are unable to classify dwarfs due to the absence of redshift data.

GALFIT and morphology based on visual inspection was determined for all non-dwarfs. In the case of dwarfs, GALFIT  and visual morphology  with a single S\'ersic and/or bulge-disk components was not possible for all galaxies.  For the analysis in this paper,  we required parameters like $B/T$, S\'ersic index of the bulge $n$ and the Kormendy relation. As we could not  determine these for dwarfs,  we restricted our sample to non-dwarfs.

\begin{table}
 \begin{tabular}{ |c|l|l|  }
 \hline
 
 Col& Parameter& Description\\
 \hline
1 & COMA\_ID &Name of source.\\
2&RA (J2000)&Right ascension of source.\\
3&Dec. (J2000)&Declination of source.\\
4 &$Bm$&Bulge magnitude\\
 5 & $Bm_{error}$&Error in bulge magnitude\\
6&$BSB\mu_e$ &Bulge  surface brightness\\
7&$Br_e$&Bulge effective radius in arcsec\\
8 & $Bn$ & Bulge S\'ersic index\\
9&B$n_{error}$&  Error in bulge S\'ersic index\\
10&$Bb/a$&Bulge axis ratio\\
11&$Bb/a_{error}$& Error in bulge axis ratio\\
12& $BPA$& Bulge position angle\\
13&$BPA_{error}$& Error in bulge position angle\\
14 &$Dm$&Disk magnitude\\
15 &$Dm_{error}$& Error in disk magnitude\\
16&$DSB\mu_e$ & Disk surface brightness\\
17&$Dr_s$&Disk scale length in arcsec\\
18&$Db/a$& Disk axis ratio\\
19&$Db/a_{error}$& Error in disk axis ratio\\
20&$DPA$ &Disk position angle\\
20&$DPA_{error}$ & Error in disk position angle\\
21&$z$&Redshift\\
22&$B/T$&Bulge to total light ratio\\
23&Morphology&Based on visual inspection\\
24 &$Gm$& Galaxy magnitude\\
25 &$Gm_{error}$& Error in Galaxy magnitude\\
26&$G\mu_e$ & Galaxy surface brightness\\
27&$Gr_e$& Galaxy effective radius in arcsec\\
28&$Gn$& Galaxy S\'ersic index\\
29&$Gn_{error}$& Error in Galaxy S\'ersic index\\
30&$Gb/a$& Galaxy axis ratio\\
31&$Gb/a_{error}$& Error in Galaxy axis ratio\\
32&$GPA$ &Galaxy position angle\\
33&$GPA_{error}$ & Error in Galaxy position angle\\
 \hline
\end{tabular}
\caption{Results of Bulge-Disk decomposition and single S\'ersic fits. }
\label{galfitall}
\end{table}

\begin{table}
 \begin{tabular}{ |p{1.5cm}||p{1.5cm}|p{1.5cm}|p{1.5cm}|  }
 \hline
 Morphological type& Member&Non-member&Unknown redshift\\
 & 51&28&55\\
 \hline
  \vspace{.05cm} E/SO &  \vspace{.05cm} 9(18\%)&  \vspace{.05cm}7(25\%)&  \vspace{.05cm} 7(13\%)\\
 cD & 1(2\%) & & \\
 E/SBO & 6(12\%) & &\\ 
 SO & 15(29\%) & 7(25\%) & 16(29\%)\\
 SBO  &11(22\%)&3(11\%)&7(13\%)\\
 Sp& 3(6\%)&7(25\%)&15(27\%)\\
SBp&3(6\%)&2(7\%)&5(9\%)\\
 Ring& 3(6\%)  & &\\
 Irr&  & 2(7\%)&5(9\%)\\
\hline
\end{tabular}
\caption{Morphological Distribution of the non-Dwarf Galaxies }
\label{morph}
\end{table}



\subsection{Bulge to Total Light Ratio ($B/T$) and Morphology Relation}

$B/T$ is related to the  morphological type of galaxies and increases progressively from late-type to early-type  \citep{1926ApJ....64..321H,2010MNRAS.409..405H}.
In this section, we investigate  how the  $B/T$ varies with morphological types for cluster members, non-members and unknown redshift galaxies. 

Galaxies can be classified according to their $B/T$ ratio. \cite{1986ApJ...302..564S} classified them as follows:  $ 0.8 < B/T $ (elliptical), $0.6<B/T \leq 0.8$ (elliptical with disk), $0.48 < B/T \leq 0.6$  (SO). Further, they divide spirals into two categories: early-type and late-type spirals. $B/T$ for  is defined as  $0.24 < B/T \leq 0.48$ early-type spirals and for late-type spirals is $0<B/T\leq 0.24$.

\cite{2007Ap&SS.312...63H} classifies galaxies based on their $B/T$  as follows:
$0.8 < B/T < 1 $ (E/SO), $0.5 < B/T < 0.8$ (SO), $0.15 < B/T < 0.5$ (Sab) and $0 <B/T < 0.15$ (Sbc).  This has been described in Table \ref{btmem1}.

\begin{table}
 \begin{tabular}{ |c|c|  }
 \hline
 \multicolumn{2}{|c|}{\cite{1986ApJ...302..564S} classification Bulge to Total light ratio} \\
  \hline
Morphological type & $B/T$ \\
\hline
 E/SO &  $0.8 < B/T < 1 $ \\
 SO & $0.5 < B/T < 0.8$ \\
 Early-type Sp &  $0.24 < B/T \leq 0.48$     \\
 Late-type Sp & $0<B/T\leq 0.24$ \\
\hline 
 \hline
 \multicolumn{2}{|c|}{\cite{2007Ap&SS.312...63H} classification Bulge to Total light ratio} \\
  \hline
Morphological type & $B/T$ \\
\hline
 Ellipticals &  $ 0.8 < B/T $ \\
 E/SO &  $0.6<B/T \leq 0.8$ \\
 SO & $0.48 < B/T \leq 0.6$ \\
Sab &  $0.15 < B/T < 0.5$     \\
Sbc & $0 <B/T < 0.15$ \\
\hline 
\end{tabular}
\caption{Classification of morphological types using $B/T$ ratio by\protect\cite{1986ApJ...302..564S} and \protect\cite{2007Ap&SS.312...63H}.}
\label{btmem1}
\end{table}

For our sample, there is a variation in the $B/T$ vs  morphological types for cluster members, non-members and unknown redshift galaxies. This parameter is distance independent hence can be used to compare all three categories. However, for  non-members  and unknown redshift galaxies (with an exception of five, two non-member galaxies and three unknown redshift galaxies) they agree with the earlier $B/T$  classification of morphological type (Table \ref{btmem}).

For Coma cluster galaxies \cite{2004ApJ...602..664G} observed that $0.3 < B/T$ for early-type galaxies (E \& SO) and $B/T<0.3$  for Sp and $0.6 < B/T$ for E. \cite{2004AJ....127.1344A} observed that $0.6 < B/T$ for E, E/SO and $B/T<0.6$ for SO \& Sp.  

Except for one galaxy in our sample of member galaxies, we have agreement with these $B/T$ ratios (Table~\ref{btmem1}).   
Based on the residual this galaxy was classified by as SBO in spite of it having a $B/T=0.9$. This classification is in agreement with \cite{2012ApJ...746..136M}.

\begin{figure}
\centering
\includegraphics[width=7cm,height=3cm]{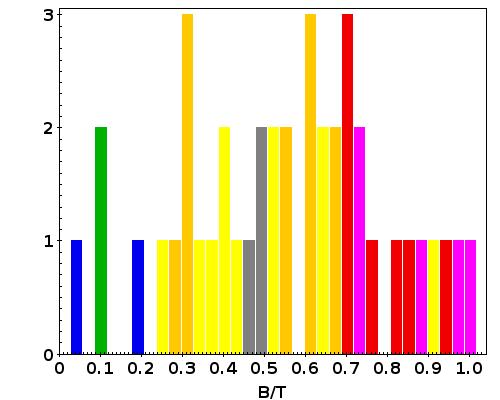}\\
\includegraphics[width=7cm,height=3cm]{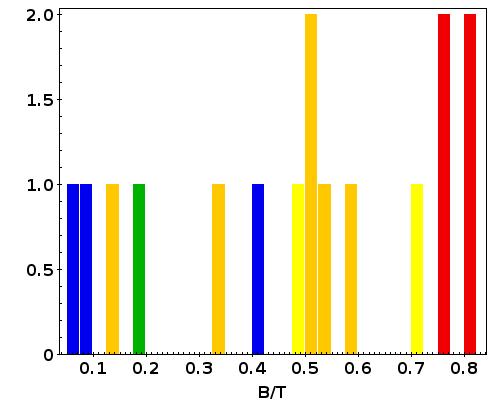}\\
\includegraphics[width=7cm,height=3cm]{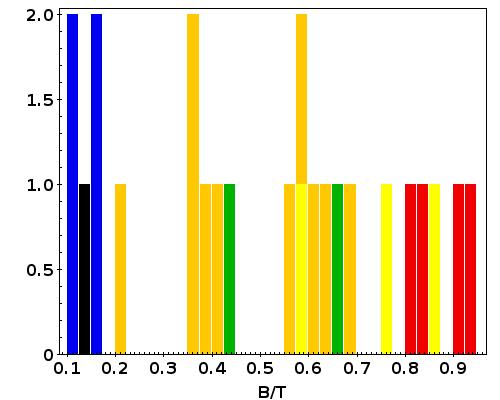}\\
\caption{Morphology distribution and $B/T$ value for cluster members, non-members and unknown redshift galaxies. In the figure,  E/SO (red), E/SBO (pink), SO (dark yellow), SBO (light yellow), Sp (blue),  SBp (green), ring (grey) and  irregular (black).}
\label{bt}
\end{figure}

In Figure \ref{bt}, we show the  morphology  distribution of $B/T$ for members, non-members and unknown redshift galaxies excluding the dwarf galaxies. 

   


As is evident from Figure \ref{bt} our results agree with those of \cite{2004ApJ...602..664G,2004AJ....127.1344A} where,  for Coma Cluster members, early-type (E \& SO) $0.3 < B/T$ and  late-types  $B/T < 0.3$. 

We observe that in general,  $B/T$ increases with morphological types from spirals to ellipticals. 

\begin{table}
 \begin{tabular}
{ |p{0.5cm}||p{2.3cm}|p{2.3cm}|p{2.3cm}|  }
  \hline
 Visual type & Member & Non-member & Unknown redshift\\
\hline
 E/SO &$0.69 \leq B/T \leq0.94$  &$0.75\leq B/T \leq 0.81$&$0.81\leq B/T \leq0.94$\\
 E/SBO& $0.72 \leq B/T \leq 0.99$&  &  \\
 SO &$0.28 \leq B/T\leq 0.67$&$0.13\leq B/T\leq 0.59$ &$0.21\leq B/T \leq0.67$\\ 
 SBO  &$0.25 \leq B/T \leq 0.63$ & $0.16 \leq B/T \leq 0.7$& $0.58\leq B/T \leq 0.86$\\
 Sp&  $B/T \leq 0.05$  & $0.05 \leq B/T \leq 0.4$&$0.11\leq B/T \leq 0.17$\\
SBp& $B/T \leq 0.11$ & $B/T \leq 0.19$   &$0.42 \leq B/T \leq 0.66$\\
 Ring& $0.45 \leq B/T \leq 0.49$  & &\\
 Irr&  & &$B/T\leq0.12 $\\
\hline 
\end{tabular}
\caption{Distribution of $B/T$ of the complete sample (excluding dwarfs).}
\label{btmem}
\end{table}


\subsection{Color Magnitude Relation (CMR)}
 CMR is an important, distance-independent parameter used to study the characteristics of galaxies \citep{1977ApJ...216..214V,1992MNRAS.254..589B}. Galaxy  morphology is related to the slope and scatter of the CMR \citep{2001MNRAS.326.1547T,2010ApJS..191..143H}. 

\begin{figure}
\centering
\includegraphics[width=8cm,height=5cm]{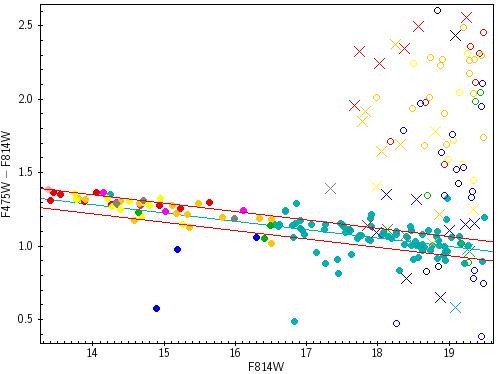}\\
\caption{CMR for cluster members, non-members and unknown redshift galaxies and their morphological types. Filled circles represents the members, cross marks represents the non-members and open circles represents the unknown redshift galaxies. The solid line is the linear fit of cluster members,  E/SO (red), E/SBO (pink), SO (dark yellow), SBO (light yellow), Sp (blue),  SBp (green), ring (grey) and  irregular (black).}
\label{cmr}
\end{figure}

\cite{2010ApJS..191..143H} noted that in the CMR, the red sequence lies in the apparent magnitude range  $ 13^m < F814W < 22.5^m$. In our non-dwarf sample,  the majority of  galaxies are E/SO, E/SBO, SO and  SBO which belong to the red sequence. The exceptions are 5 galaxies, which are Sp (3) and SBp (2) and these lie below the red sequence. Figure \ref{cmr} shows the CMR of the cluster members, non-members and unknown redshift galaxies as a function of morphological type. Colors  ($F814W-F475W$) have been calculated from the SExtractor  catalog. Extinction corrections have been made using values from \cite{2010ApJS..191..143H}.  We fit a solid line of the member galaxies (excluding the SBp and Sp) and obtain the relation $$F475W-F814W=(-0.057 \pm 0.00558)*M_{F814W}+(0.208 \pm 0.09429)$$ 
 A similar fit was found by \citep{1992MNRAS.254..589B,2001MNRAS.326.1547T}. Members are within  $1\sigma = 0.057^m$ of the solid line. 

We observe that as the magnitude decreases,   the color of the galaxies gets bluer.  Spiral galaxies are  brighter and bluer due to their high star formation rates (SFR). This is probably due to Star Formation (SF)  induced by cluster tidal effects or high speed galaxy-galaxy encounters and is  related to environment of cluster. \citep{1998ApJ...509..587F}.  The SBp galaxies (green) are closer to the red sequence compared to the Sp. The color variation in galaxies can be explained by differences in metallicity using stellar population models from \cite{1996ApJS..106..307V,1994ApJS...95..107W}.  

 \subsubsection{Relation between the color of galaxies and $B/T$}
\begin{figure}
\centering
\includegraphics[width=8cm,height=5cm]{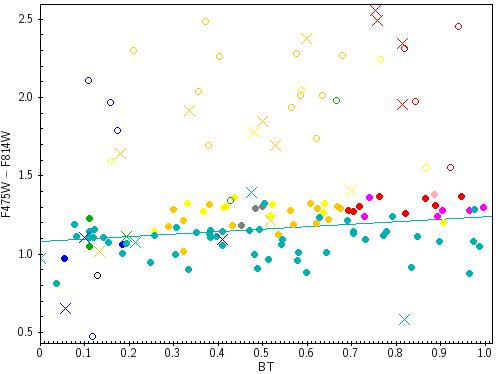}\\
\caption{Relation between color and $B/T$ for cluster members (filled circles), nonmembers (cross marks) and unknown redshift (open circles) galaxies with morphological types,E/SO (red), E/SBO (pink), SO (dark yellow), SBO (light yellow), Sp (blue),  SBp (green), ring (grey) and  irregular (black).}
\label{cbt}
\end{figure}

 In this section, we explore the color dependence of $B/T$ as a function of morphological type.
 Figure~\ref{cbt}  shows the relation between color and $B/T$ for  cluster members, non-members and unknown redshift galaxies. We observe that the color of galaxies gets redder with increasing $B/T$  except for dwarfs where the scatter is large. 
 
 Both color and $B/T$ are distance independent and hence can be plotted for the complete sample. Dwarfs cover a range of $B/T$, but colors are limited to $(F475W -F814W) < 1.2^m$ (including spirals) for members. 
Ellipticals are redder than dwarfs and the color $(F457W - F814W)>1.2^m$. Spirals (Sp) have $(F475W − F814W ) \leq 1^m$. One SBp has similar colors to SBO and the other is similar to Sp.
Though we have only 3 Sp and 2 SBp galaxies, we observe that SBps have a redder color value compared to Sp. 

In general, we observe that spirals are in the lower (blue) region, ellipticals are in upper (red) region and lenticulars (SO \& SBO) are in between indicating a possible evolutionary sequence.
In the case of unknown redshift galaxies, a few objects classified as Sp and SBp have high $B/T$ and behave like early-type galaxies in the CMR diagram. The GALFIT for these objects has larger errors which could explain the $B/T$ variation. But the red colors indicate low SFR probably due to quenching. 
 
\subsection{S\'ersic index}

 S\'ersic index $n$ also plays an important role in galaxy classification and is distance independent. The S\'ersic index of the bulge is used to classify the bulge as classical ($n> 2$) or pseudo ($n<2$)  \citep{2008AJ....136..773F}. There can  be a 20\% ($\pm 0.5$) error in the determination of this parameter \cite{2009MNRAS.393.1531G}.
 
 The S\'ersic index of the galaxy can  also be used to classify the galaxies as  dwarfs  ($n<2$) and giants ($n>2$) \citep{2004ApJ...602..664G}. \cite{2005ApJ...635..959B, 2008ApJ...675L..13V}  found through simulations that early-type galaxies have $n>2.5$ and late-type have $n < 2.5$.  
 
\begin{figure}
\begin{tabular}{ccc}
\centering
\includegraphics[width=2.5cm,height=3cm]{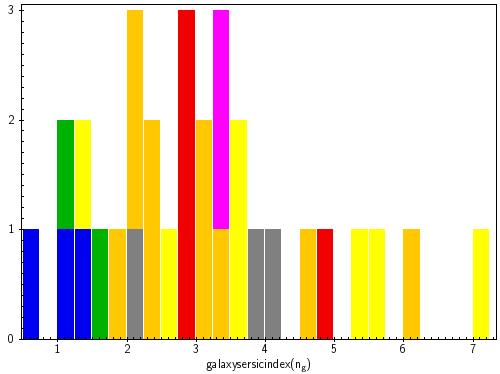}&
\includegraphics[width=2.5cm,height=3cm]{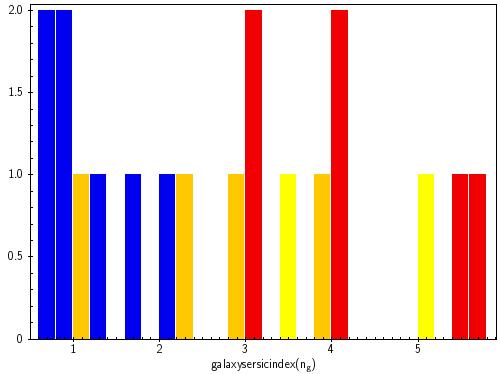}&
\includegraphics[width=2.5cm,height=3cm]{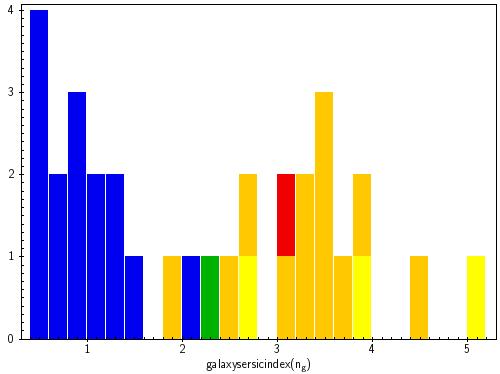}\\
\includegraphics[width=2.5cm,height=3cm]{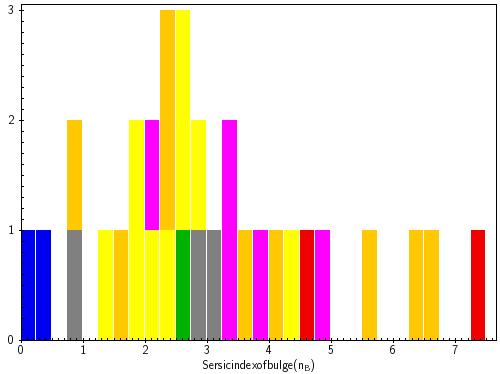}&
\includegraphics[width=2.5cm,height=3cm]{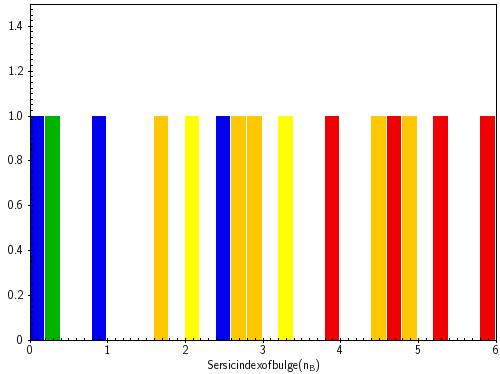}&
\includegraphics[width=2.5cm,height=3cm]{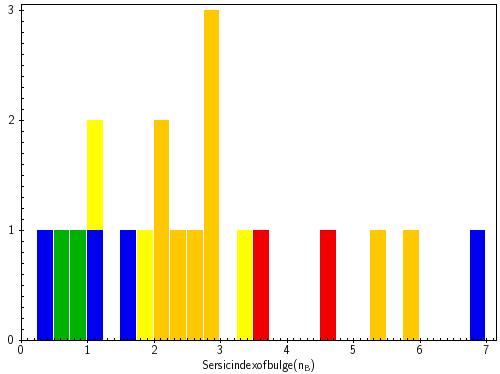}\\
\end{tabular}
\caption{The Morphology distribution and S\'ersic  index of the galaxy and its bulge for cluster members, non-members and unknown redshift galaxies. E/SO (red), E/SBO (pink), SO (dark yellow), SBO (light yellow), Sp (blue),  SBp (green), ring (grey) and  irregular (black).}
\label{snn}
\end{figure}

 Figure \ref{snn} shows the  distribution of the bulge and galaxy S\'ersic indices for  cluster members, non-members and unknown redshift galaxies.
 
 The S\'ersic index of the galaxies agrees with  previous studies where early-type have $n> 2.5$ and late-type $n< 2.5$  for our complete sample of 219 objects (members, non-members and unknown redshift galaxies) and is shown in Fig. \ref{snn}. In the case of SOs, the fainter galaxies have $n < 2$ and the brighter ones have $n>2$ which agrees with  \cite{2013ApJ...767L..33V}.
 
In the case of dwarf galaxies which  have $n < 2$,  we have 9 member dwarfs with $n > 2$. These could possibly be dwarfs with nuclear star clusters  and ultra  compact galaxies \citep{2008ApJ...674..653E, 2009MNRAS.397.1816P, 2014MNRAS.445.2385D}.

\begin{figure}
\centering
\includegraphics[width=8cm,height=3cm]{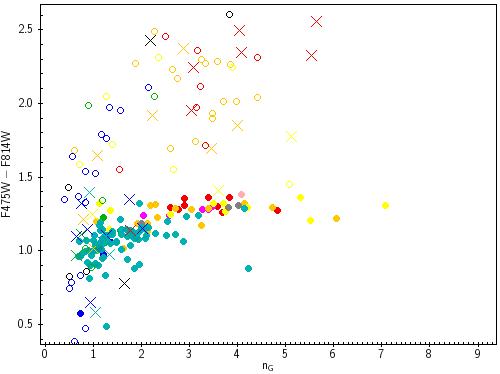}\\
\includegraphics[width=8cm,height=3cm]{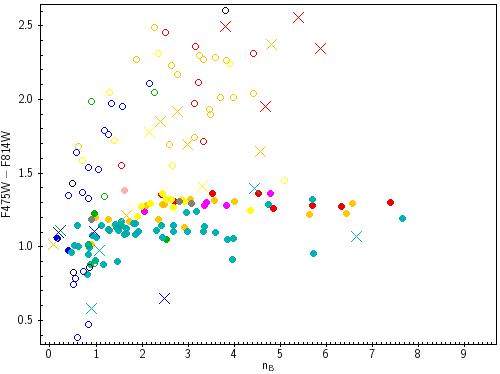}\\
\caption{Relation between color and S\'ersic index of  galaxy and bulge  for cluster members (filled circles), nonmembers (cross marks) and unknown redshift (open circles)  galaxies with morphological types. E/SO (red), E/SBO (pink), SO (dark yellow), SBO (light yellow), Sp (blue),  SBp (green), ring (grey) and  irregular (black).}
\label{sgb}
\end{figure}

\begin{figure}
\centering
\includegraphics[width=8cm,height=3cm]{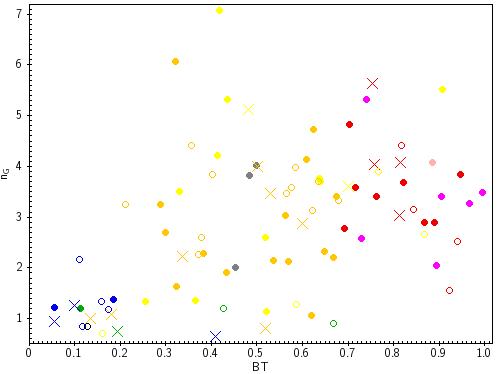}\\
\includegraphics[width=8cm,height=3cm]{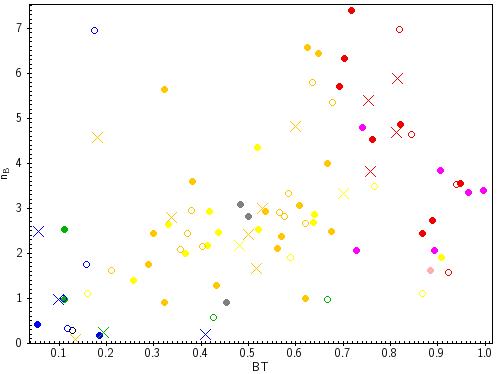}\\
\caption{Relation between  S\'ersic index  of  galaxy  and bulge and $B/T$ for cluster members (filled circles), nonmembers (cross marks) and unknown redshift (open circles)  galaxies with morphological types. E/SO (red), E/SBO (pink), SO (dark yellow), SBO (light yellow), Sp (blue),  SBp (green), ring (grey) and  irregular (black)}
\label{sbt}
\end{figure}

We shall now explore the relation between S\'ersic index (galaxy and bulge) with $B/T$ and the color of galaxies. $B/T$ and the shape parameter of the bulge is crucial in estimating how bulges were formed in galaxies: in  galaxy mergers in the hierarchical clustering, or by more slow secular evolutionary processes in galaxies.  Figure  \ref{sgb} and \ref{sbt} show the relation between S\'ersic index and $B/T$ and clearly galaxies with larger $B/T$ have larger S\'ersic index implying a formation due to mergers.  The $B/T$ is also a key parameter for evaluating the importance of gas stripping in galaxies, a process by which spirals might be converted into S0s \citep{laurikainen_salo_buta_knapen_speltincx_block_2006}.

These three parameters are independent of distance. In general,  S\'ersic index increases with $B/T$ and color gets redder with increasing $B/T$. 
 
\subsection{Kormendy relation}
The Kormendy relation \citep{1977ApJ...218..333K} is the relation between $\mu_{e}$ and $r_{e}$, where $\mu_{e}$ is the surface brightness of the galaxy at the effective radius $r_{e}$ and can be used to classify bulges and hence infer the merger history of galaxies. \cite{1987IAUS..127..379H} did a linear fit of these parameters for  elliptical galaxies. \cite{2009MNRAS.393.1531G}  found that classical bulges occupy the same area as elliptical galaxies  in the Kormendy relation. They also found that pseudo bulges lie outside the $3\sigma$ line in the Kormendy relation.

\begin{figure}
\centering
\includegraphics[width=8cm,height=5cm]{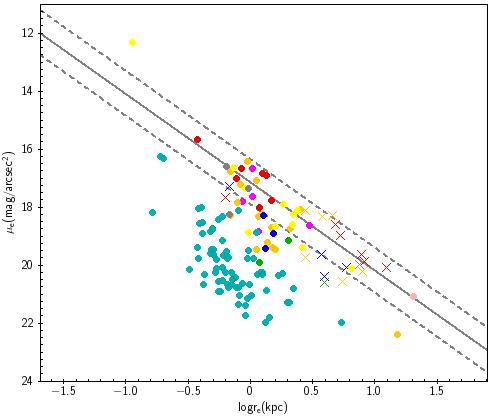}\\
\caption{Kormendy relation between surface brightness of galaxy $\mu_{e}$ at $r_{e}$ and effective radius of galaxy $ r_{e}(kpc)$ for cluster members (filled circles), nonmembers (cross marks).  E/SO (red), E/SBO (pink), SO (dark yellow), SBO (light yellow), Sp (blue),  SBp (green), ring (grey) and  irregular (black). Middle line represents the linear fit of ESO and upper \& lower lines are $\pm3\sigma$ lines.} 
\label{korgal}
\end{figure}

\begin{figure}
\centering
\includegraphics[width=8cm,height=5cm]{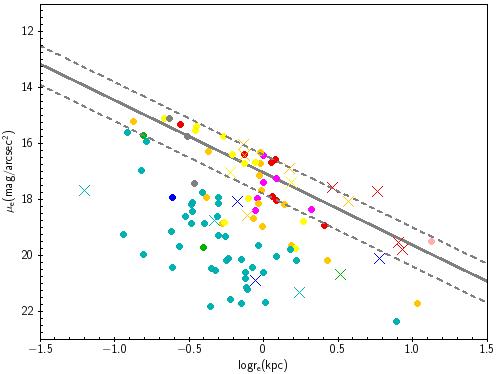}\\
\caption{Kormendy relation between surface brightness of bulge $\mu_{e} $ at $r_{e}$ and effective radius of bulge $ r_{e}(kpc)$ for cluster members (filled circles), nonmembers (cross marks). E/SO (red), E/SBO (pink), SO (dark yellow), SBO (light yellow), Sp (blue),  SBp (green), ring (grey) and  irregular (black). Middle line represents the linear fit of E/SO and upper\& lower lines are $\pm3\sigma$ lines.} 
\label{korbul}
\end{figure}

We did a linear fit of E/SO galaxies and their bulges and  found a slope  $3.04 \pm0.398 $ for galaxies and $2.604 \pm0.544$ for the bulge parameters of E/SO.  The lower and upper lines are  the $\pm3\sigma$ limits. Figures \ref{korgal} and \ref{korbul} show the Kormendy relation obtained from the S\'ersic fit of the galaxy and from the S\'ersic fit of the bulge component of the galaxy for different morphological types.
  The  galaxy fits are  similar to \cite{1977ApJ...218..333K, 1987IAUS..127..379H}
  and the  bulge fits are similar to \cite{2009MNRAS.393.1531G, 2013ApJ...767L..33V}.  According to \cite{2009MNRAS.393.1531G} the points below the $3\sigma$ line are pseudo bulges, that implies secular evolution of spirals and dwarfs which lie in that region. In both the figures we can observe that, this relation clearly agrees with our visual classification of galaxies for cluster members and non-members. The  spirals (Sp \& SBp) and dwarf galaxies are located  below the $3\sigma$ line which implies that they have a pseudo bulges. We can also see that the E/SO galaxies with classical bulges  lie with the  $3\sigma$ lines of our linear fit as described by \citep{2009MNRAS.393.1531G, 2004ARA&A..42..603K}.

\subsection{Statistical Analysis}
We studied the data obtained in Table \ref{morph} and binned the galaxies in two categories:
\begin{enumerate}
    \item E, SO, Irr, Ring
    \item Sp, SBp  
\end{enumerate}
As shown in the Table \ref{binmorph}

\begin{table}
 \begin{tabular}{ |p{1.5cm}||p{1.5cm}|p{1.5cm}|p{1.5cm}|  }
 \hline
 Morphological type& Member& Non-member& Unknown Redshift\\
 \hline
Category i & 45 & 19 &  35 \\
Category ii & 6 & 9 &  20 \\
\hline
{\bf Total}  & {\bf 51} & {\bf 28} &{\bf  55} \\
\hline 
\end{tabular}
\caption{Morphological Binned Distribution of the non-Dwarf Galaxies }
\label{binmorph}
\end{table}
We tested the following hypotheses:
\begin{itemize}
\item Hypothesis 1: Are there significantly larger number of Category~(i) galaxies in members than in non-members?\\
As per our data,  in Category (i) we have members (45), non-members (19) and unknown redshift (64). Category (ii) has members (6), non-members (9), unknown red-shifts (15). Fisher's exact test, gave $p = 0.037$. For the $\chi^2$ test, we used the   $\chi^2$ statistic $= 4.8799$. The $p$-value is 0.027171. 

The result is significant both with Fisher's and  $\chi^2$ tests which confirms our hypothesis and understanding that this cluster has a majority of E \& SOs. This means that there is enough evidence to say that there is a difference in the distributions.							
\item Hypothesis 2: Are there significantly larger number of E, SOs members than in the  non-members and unknown redshift galaxies? \\	As per our data,  in Category (i) we have members (45), non-members (19) and unknown redshift (35). Category (ii) has members (6), non-members(9), unknown red-shifts(20). For the $\chi^2$ test, we used the   $\chi^2$ statistic $= 8.79$. The $p$-value is $0.003025$. 	This implies that our hypothesis is correct. We repeated this with assuming our unknown redshift galaxies to be non-members and found that our result does not change.

\item Hypothesis 3: How does Hypothesis 2 change if we assume that our unknown redshift galaxies are members?
The $\chi^2$  statistic is $0.6655$. The $p$-value is $0.4146$. This implies	that the unknown redshift galaxies are are more likely to be non-members.
				
\item				
									
Question: Which population are the unknown redshift galaxies similar to?\\
The $\chi^2$ statistic is $0.1454$. The $p$-value is $0.70296$. This implies that unknown redshift galaxies are more likely non-members than members.
\end{itemize}

\section{Conclusions}
We have determined the morphology and structural properties of 219 galaxies from HST/ACS  data of the central region of the Coma Cluster brighter with  $F814W< 19.5^m$. To study the structural properties of the galaxies,  we used GALFIT and did a a single S\'ersic fit as well as a two dimensional bulge-disk decomposition.

We divided our sample into cluster members, non-members   and unknown redshift galaxies for which we did not have redshifts. Using statistical techniques, we find that our unknown redshift galaxies are most probably non-members and hence do not affect our analysis. 
With the help of  \cite{2012ApJ...746..136M},  we separated our sample into dwarfs and non-dwarfs, by putting a cut-off at absolute $ F814W < -18.5^m$.

We assign morphological types by inspecting the GALFIT residuals that we get from the difference of the science image and the model images for our non-dwarf sample. We found of the  132 cluster members,  51 are non-dwarfs. For the non-Dwarfs, we have cD-1 (2\%), E/SO-9 (18\%), E/SBO-6 (12\%), SO-15 (29\%), Ring-3 (6\%), SBO-11 (22\%), Sp-3 (6\%) , SBp-3 (6\%).  
 For the 32 nonmembers, 28 were non-dwarfs E/SO-7 (25\%), SBO-3 (11\%), SO-7 (25\%), SBp-2 (7\%), Sp-7 (25\%), Irr-2 (7\%). 
 For 55 unknown redshift galaxies are  E/SO-7 (13\%), SBO-7 (13\%), SO-16 (29\%), SBp-5 (9\%), Sp-15 (27\%) and Irr-5 (9\%).      
We used this classification to compare and plot  the structural properties of galaxies, $B/T$ ratio, Kormendy relation, CMR and  their correlations for cluster members, non-members and unknown redshift galaxies.

 We compared $B/T$ ranges to our visual morphology with earlier works. The limits of $B/T$ vary for members, non-members and unknown redshift galaxies. Coma members are in agreement with previous studies \citep{2004ApJ...602..664G, 2004AJ....127.1344A} where early-type galaxies (E/SO, SO and SBO) and late-type galaxies (Sp \& SBp) have a separation at $B/T=0.3$.  In members, only one object that we classified as SBO has $B/T=0.9$. This classification is in agreement with \cite{2012ApJ...746..136M}.  
 We  have also compared  $B/T$ with non-members and unknown redshift galaxies. Our morphologies with $B/T$  agree with \cite{1986ApJ...302..564S} and \cite{2007Ap&SS.312...63H}, except for two galaxies that are non-members  and three galaxies that are unknown redshift galaxies. These might be because the two objects  (non-members) are edge-on galaxies. 
 
We find that non-members \& unknown redshift galaxies have $B/T$ values similar to field galaxies compared  to the member galaxies.

We have also plotted the CMR for different morphological types in our sample. For the cluster members our results are in agreement with \cite{2010ApJS..191..143H}. For members, E/SO, E/SBO, SO, SBO and Ring galaxies are in red sequence while Sp and SBp  are in the blue region. These are much brighter and  have a larger scatter from the linear fit due to their SFRs. 
 We find that the color of the galaxy increases with $B/T$. 
In the case of unknown redshift galaxies, a few objects classified  as Sp and SBp have high $B/T$ and behave like early-type galaxies in the CMR diagram. The GALFIT for these objects have larger errors which could  lead to incorrect $B/T$s. However, these galaxies have redder colors which indicate low SFR,  probably due to quenching.
In  general, we observe that spirals are in the lower (blue) region, ellipticals are in upper (red) region and lenticulars (SO \& SBO) are intermediate region, indicating a possible evolutionary sequence.

We found that the S\'ersic index of the galaxy and bulge are in agreement with values for morphological types found in previous studies  for early-type and late-type galaxies. The only exceptions are 9 dwarf members. These 9 objects might be ultra compact dwarf galaxies (UCD), nucleated dwarfs (dE N) or non-nucleated dwarf (dE NN) galaxies  \citep{2008ApJ...674..653E,2009MNRAS.397.1816P,2014MNRAS.445.2385D}. They lie in the same region UCDs, dE N  and dE NN in the  CMR of \cite{2010ApJS..191..143H}.  

 We plotted the Kormendy relation for our sample and found that the early-type galaxies with classical bulges lie within the $3\sigma$ range of the linear fits, while the late-type galaxies and dwarfs lie in the lower region away from the linear fit. 
 
 The present study shows that the majority of member galaxies  in the core of Coma Cluster are E, E/SOs (classical bulges) and dwarfs (psuedo bulges) as shown by our statistical study in the earlier section. These are results of mergers as the majority (75\%) of members have classical bulges, while only 25\% have psuedo bulges. Coma Cluster has an age of 10 Gyr and there is a clear absence of member spiral and interacting galaxies, which indicate that interactions may have been fast and at later stages as member galaxies have high velocity dispersions of 1000 km/s \citep{2003astro.ph..5512M}. In future papers, we shall further explore the E/SO population with an emphasis on their merger history.




\section*{Acknowledgements}
 The authors would like to thank the referee for very useful and important suggestions to improve the paper.
 
This paper is based on observations made with the NASA/ESA Hubble Space Telescope, obtained from the data archive at the Space Telescope Science Institute. STScI is operated by the Association of Universities for Research in Astronomy, Inc. under NASA contract NAS $5-26555$.


\section*{Data Availability}

 The data underlying this article (Table~\ref{galfitall}, Table~\ref{terr}, Table~\ref{singleser} and Table~\ref{bdc}) will be shared on request to the correspondg author  \footnote{The table \ref{galfitall} is available at \url{https://drive.google.com/file/d/1JT5io1EFdS8W_nnua4pNRXwUp174BLQ6/view?usp=sharing }}



\bibliographystyle{mn}
\bibliography{myref}

\appendix

\section{Supplementary Tables}

Table \ref{terr}1 presents the errors in parameters obtained from  GALFIT for  single S\'ersic fits and bulge-disk decomposition. 
\begin{table}
 \begin{tabular}{ |c|l|l|  }
 \hline

 Col& Parameter& Description\\
 \hline
1 & COMA\_ID &Name of source.\\
2&RA (J2000)&Right ascension of source.\\
3&Dec. (J2000)&Declination of source.\\
4 & $Bm_{error}$& Error in bulge magnitude\\
5&$Bn_{error}$& Error in bulge S\'ersic index\\
6&$Bb/a_{error}$&Error in bulge axis ratio\\
7&$BPA_{error}$& Error in bulge position angle\\
8 &$Dm_{error}$&  Error in disk magnitude\\
9&$Db/a_{error}$&  Error in disk axis ratio\\
10&$DPA_{error}$ & Error in disk position angle\\
11 &$Gm_{error}$& Error in Galaxy magnitude\\
12&$Gn_{error}$& Error in Galaxy S\'ersic index\\
13&$Gb/a_{error}$& Error in Galaxy axis ratio\\
14&$GPA_{error}$ & Error in Galaxy position angle\\
 \hline
\end{tabular}
\caption{Errors in the parameters obtained from  GALFIT for  Bulge-Disk decomposition and single S\'ersic fits.}
\label{terr}
\end{table}

Table \ref {singleser}2 shows a comparison of the single Sersic fits obtained by this work and that of \cite{2011MNRAS.411.2439H,2014MNRAS.441.3083W}. 
In the tables, to  facilitate comparison of parameters, parameters like $r_e$ and $r_s$  were converted to arcsec using the average distance of Coma Cluster (100~Mpc) and the pixel scale. We also converted  absolute to apparent magnitudes using distance modulus. The parameter $b/a$ was converted to  ellipticity from the axis ratio using $b/a=1-e$.
 
 \begin{table}
 \begin{tabular}{|llll|}
 \hline
 \hline
\multicolumn{1}{|l|}{This work}&
\multicolumn{1}{|l|}{\cite{2011MNRAS.411.2439H}}&
\multicolumn{1}{|l|}{\cite{2014MNRAS.441.3083W}}\\

\hline
  Parameter (Col) &Parameter (Col)&Parameter (Col)& Description\\
 \hline
 COMA\_ID (1) && &Name of source.\\
RA (J2000)(2)&&&Right ascension of source.\\
Dec. (J2000)(3)&&&Declination of source.\\

$Gm$(4)&$Gm\_1$(10)&$Sm$(16)& Galaxy magnitude\\
$G\mu_e$(5) & $G\mu_e\_1\ (11)$&&Galaxy surface brightness\\
$Gr_e$(6)&$Gr_e\_1$(12)&$Sr_e$\ (17)& Galaxy effective radius in arcsec\\
$Gn$(7)&$Gn\_1$(13)&$Sn$\ (18)& Galaxy S\'ersic index\\
$Gb/a$\ (8)& $Gb/a\_1$\ (14)&& Galaxy axis ratio\\
$GPA$\ (9) & $GPA\_1$\ (15) & &Galaxy position angle\\

 \hline
\end{tabular}

\caption{Comparison of the single S\'ersic fits obtained by this work with \protect \cite{2011MNRAS.411.2439H} and \protect \cite{2014MNRAS.441.3083W}. In the table,  $G$ parameters are from this work, all $G\_1$ are from \protect \cite{2011MNRAS.411.2439H} and $S$ parameters are from \protect \cite{2014MNRAS.441.3083W}. }
\label{singleser}
\end{table}

Table \ref{bdc}3 shows the comparison of the bulge-disc decomposition  obtained by this work and that of  \cite{2014MNRAS.440.1690H,2014MNRAS.441.3083W}. 

\begin{table}
 \begin{tabular}{|l|l|l|l|}
 
 \hline
 \hline
\multicolumn{1}{|l|}{This work}&
\multicolumn{1}{|l|}{\cite{2014MNRAS.440.1690H}}&
\multicolumn{1}{|l|}{\cite{2014MNRAS.441.3083W}}\\

\hline
  Parameter (Col) &Parameter (Col)&Parameter (Col)& Description\\
 \hline
 COMA\_ID (1) && &Name of source.\\
RA (J2000)(2)&&&Right ascension of source.\\
Dec. (J2000)(3)&&&Declination of source.\\

$Bm$(4)&$Bm\_1$(15)&& Galaxy magnitude\\
$B\mu_e(5)$& & & Bulge surface brightness\\
$Br_e$ (6)&$Br_e\_1$ (16) & $S1r_e$(24) & Bulge effective radius in arcsec\\
$Bn$(7) & $Bn\_1$(17)&$S1n$(25)& Bulge S\'ersic index \\
$Bb/a$(8)&$Bb/a\_1$(18)& & Bulge axis ratio\\
$BPA$(9)&$BPA\_1$(19)& & Bulge position angle\\
$Dm$ (10)& $Dm\_1$ (20)& &Disk magnitude\\
$D\mu_e$\ (11)& & & Disk surface brightness\\
$Dr_s$\ (12)& $Dr_s\_1$(21)& $S2r_e$(26) & Disk scale length in arcsec\\
$Db/a$\ (13) & $Db/a\_1$\ (22) & &Disk axis ratio\\
$DPA$\ (14) &$DPA\_1$\ (23) & & Disk position angle\\

\hline
\end{tabular}
\caption{Comparison of results of Bulge-Disk of this work with \protect \cite{2014MNRAS.440.1690H} (Bulge+Disk decomposition) and \protect \cite{2014MNRAS.441.3083W} (Multiple S\'ersic). Here all $B$ and $D$ parameters are from this work, all $B$\_1 and $D$\_1 are from \protect \cite{2014MNRAS.440.1690H} and S1 and S2 parameters are from \protect \cite{2014MNRAS.441.3083W}.}
\label{bdc}

\end{table}


\bsp	
\label{lastpage}
\end{document}